\documentclass[aps,prl,preprint,groupedaddress,showpacs]{revtex4}

\usepackage{graphicx}
\usepackage{dcolumn}
\usepackage{bm}

\begin{document}


\title{Evidence of double magicity of $N=Z$ nuclei near the rp-process path}


\author{M.M. Sharma}
\email[]{sharma@kuc01.kuniv.edu.kw}
\affiliation{Physics Department, Kuwait University, Kuwait 13060, Kuwait}
\author{J.K. Sharma}
\affiliation{Physics Department, St. John's College, Agra-282002, India}

\date{\today}

\begin{abstract}
$N=Z$ nuclei above Ni are understood to be waiting-point nuclei in 
the rp-process nucleosynthesis. Investigating the experimental isotope 
shifts in Kr isotopes near the proton drip-line, we have 
discovered that $N=Z$ rp-process nuclei $^{68}$Se, $^{72}$Kr, $^{76}$Sr 
and $^{80}$Zr exhibit a significant shell gap both at the proton and 
neutron numbers in the deformed space with the consequence that pairing 
correlations for protons and neutrons vanish, thus lending a 
double-magic character to these nuclei. A significant number of nuclei 
in this region are also shown to exhibit neutron magicity at 
$N=$ 34, 36, 38, and 40 in the deformed space.
\end{abstract}


\maketitle


In rapid-proton-capture (rp-process) in x-ray bursters, $N=Z$ nuclei above Ni 
($Z=28$) are understood to play an important role in producing a part of x-ray 
flux \cite{Schatz.98,Schatz.01,Wiescher.01}. A number of these
nuclei are considered to be waiting-point due to inability of these nuclei 
to capture a proton, thus hindering the  rp-process. Consequently, peaks in 
abundances of the $N=Z$ nuclei $^{68}$Se, $^{72}$Kr, $^{76}$Sr and $^{80}$Zr 
among others have been shown to arise in astrophysical rp-process in x-ray 
bursters \cite{Schatz.01}.  

Precise measurements \cite{Clark.04,Rodriguez.04,Clark.07,Gomez.08} of nuclear 
masses have established the waiting-point character of the $N=Z$ nuclei 
$^{68}$Se, $^{72}$Kr, $^{76}$Sr, and $^{80}$Zr. On $^{64}$Ge, however, there
is no consensus as yet whether it is a waiting-point nucleus 
\cite{Schury.07}. Negative values of $Q_p$ for a proton capture are 
cited as evidence for the waiting-point character of nuclei.
This is interesting, since no known magic number is present in this region 
in contrast to waiting-point nuclei arising due to the major magic numbers 
along the r-process path. This poses a theoretical challenge to 
understand the structure of the waiting-point nature of rp-process nuclei 
responsible for x-ray bursts. 

The structure of these nuclei is far from certain. What is known is that 
nuclei in this region are highly deformed 
\cite{Petrovici.96,Sarri.05,Langanke.03} until the major magic number 
approaches at $Z=N=50$ where the rp-process is predicted to 
terminate \cite{Schatz.01}. In this letter, we examine the experimental 
isotope shifts in proton-rich Kr nuclei near the rp-process path 
within the framework of the relativistic Hartree-Bogoliubov 
theory in deformed space. Consequently, we report on the double-magic 
structure that we have discerned for $N=Z$ nuclei  near the rp-process path.

\begin{figure}[h*]
\resizebox{0.65\textwidth}{!}{%
  \rotatebox{0}{\includegraphics{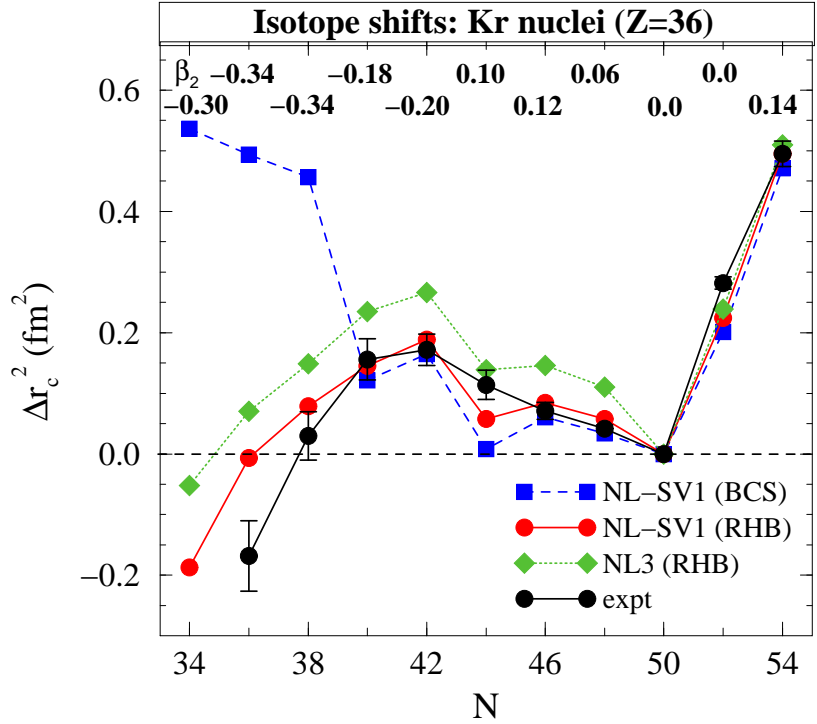}}
}
\caption{The isotope shifts $\Delta r_c^2$ of Kr chain obtained
from deformed RMF+BCS and RHB calculations with NL-SV1.
Deformation $\beta_2$ for the ground-state of nuclei obtained 
with NL-SV1 (RHB) are shown at the top of the figure. The experimental 
data \cite{Keim.95} for $\Delta r_c^2$ are shown for comparison. 
The results with NL3 are also shown. 
\label{fig1}}
\end{figure}

Isotope shifts in nuclei reveal shell effects across a major shell gap
\cite{Otten.89}, for instance, for the Pb isotopes in terms of a kink about 
the magic number \cite{Sharma.93}. For other isotopic chains such 
as Kr and Sr, such a behavior is shrouded by deformation of nuclei. 
In this work, we investigate the isotope shifts $\Delta r_c^2$  in Kr 
nuclei. The experimental values (with $N=50$ as a reference point) 
for the chain are known for isotopes down to $N=36$ with a significant 
precision from laser spectroscopy measurements \cite{Keim.95}
(see Fig~\ref{fig1}). The nuclide $^{72}$Kr ($N=36$) is very close to the 
proton drip line. The salient feature of these data (Fig.~1) 
is the monotonous increase
of $\Delta r_c^2$ from $N=50$ down to $N=40$ due to an increasing
deformation in going to the lighter isotopes. A similar feature is seen
for isotopes heavier than $N=50$. However, the most interesting feature 
that is demonstrated by the experimental data is that $\Delta r_c^2$
for $^{74}$Kr is nearly zero and that it becomes strongly negative for 
$^{72}$Kr. The magnitude of the isotope shift
or in other words, the charge radius of a nucleus with respect 
to that of the reference nucleus is mainly reflective of the deformation 
of a nucleus.  To the first order, it can be approximated as
\begin{equation}
\Delta r_c^2  = 
\langle r_c^2\rangle_s\frac{5}{4\pi} \delta \beta_2^2,
\end{equation}
where $\langle r_c^2\rangle_s$ is the mean-square charge radius 
of the spherical nucleus and $\beta_2$ is the quadrupole deformation 
of the deformed one. This excludes any effects due to shell structure. 
The downward trend of the experimental isotope shift below $N=40$ whilst
retaining $\beta_2$ in the same range would imply a structural factor. Here, 
we investigate this factor that leads to a decreasing or even a negative
isotope shift.

We examine the isotope shifts of Kr nuclei within the framework of the
relativistic mean-field theory. In this work, we employ the standard 
RMF Lagrangian with the exchange of $\sigma$, $\omega$ and $\rho$ mesons
between the nucleons. The corresponding Lagrangian density which 
describes the nucleons as Dirac spinors moving in meson fields is given by
\begin{eqnarray}
{\cal L}&=& \bar\psi \left( \rlap{/}p - g_\omega\rlap{/}\omega -
g_\rho\rlap{/}\vec\rho\vec\tau - \frac{1}{2}e(1 - \tau_3)\rlap{\,/}A -
g_\sigma\sigma - M_N\right)\psi\nonumber\\
&&+\frac{1}{2}\partial_\mu\sigma\partial^\mu\sigma-U(\sigma)
-\frac{1}{4}\Omega_{\mu\nu}\Omega^{\mu\nu}+ \frac{1}{2}
m^2_\omega\omega_\mu\omega^\mu\\ &&+\frac{1}{2}g_4(\omega_\mu\omega^\mu)^2
-\frac{1}{4}\vec R_{\mu\nu}\vec R^{\mu\nu}+
\frac{1}{2} m^2_\rho\vec\rho_\mu\vec\rho^\mu -\frac{1}{4}F_{\mu\nu}F^{\mu\nu}.
\nonumber
\end{eqnarray}
$U(\sigma)$ represents the conventional nonlinear $\sigma$ potential.
This Lagrangian includes the vector self-coupling of $\omega$-meson
represented by the coupling constant $g_4$. 
In this work, we employ the Lagrangian set NL-SV1
with the vector self-coupling of $\omega$ meson. The force NL-SV1 
was developed with a view to soften the high-density equation of state of
nuclear matter and has been shown to improve the shell effects of nuclei 
along the stability line \cite{SFM.00}. 

Deformed RMF+BCS calculations have been performed using an expansion of
fermionic and bosonic wavefunctions in 20 oscillator shells. 
The pairing is included using the constant gap 
approximation with pairing gap being taken from 
the widely used prescription of $\Delta_{n(p)} = 4.8 N^{-1/3}(Z^{-1/3})$
for open-shell nuclei \cite{Moeller.92}. The results of RMF+BCS 
calculations  of charge radii 
obtained with NL-SV1 are shown in Fig.~1. The isotope shifts
thus obtained show an increasing trend in going below $N=50$ except for
$N=44$ where the experimental data is underestimated by the theory due to 
a relatively smaller deformation $\beta_2 \sim 0.10$ as compared to its 
neighbors. The experimental values are reproduced satisfactorily down 
to $N=40$ as well as above $N=50$. However, for the isotopes 
$^{74}$Kr ($N=38$) and $^{72}$Kr ($N=36$), there is a strong divergence of 
the theoretical values from the experimental data, though $\beta_2$ of 
these nuclei is significantly larger than that for $^{76}$Kr ($N=40$). 
The isotope shift for $^{70}$Kr ($N=34$) shows a 
further increase. Experimental datum does not exist for this isotope.

Theoretically, the rapid increase in the charge radius of $^{74}$Kr,  
$^{72}$Kr and $^{70}$Kr appears naturally as a consequence 
of the vicinity to the proton drip line. 
The BCS smearing of occupation probabilities 
across the Fermi surface which adjoins the continuum leads to a 
swelling of the charge radius. Evidently, a BCS description of nuclei 
close to the proton drip line runs contrary to the experimental data. 
The isotope shifts of Kr and Sr isotopes were investigated in a 
previous work \cite{Lala.95} within the RMF+BCS formalism. Calculations 
performed with 12 oscillator shells gave rise to an apparently 
good agreement with the data \cite{Lala.95}. This agreement is, however, 
fortuitous, for 12 shells do not suffice to encompass the configuration space 
and hence underestimate the charge radii of neutron-deficient nuclei 
significantly.

Notwithstanding the stark disagreement of the RMF+BCS calculations with the
experimental isotope shifts, we have undertaken an inclusion of the
pairing mechanism based upon Bogoliubov quasi-particle scheme in a 
deformed basis with a view to take into account the effects due to 
coupling of the Fermi surface to the continuum. The eigenstates are
then eigenvectors of the generalized single-particle Hamiltonian 
containing the self-consistent mean-field and a pairing field 
$\hat{\Delta}$ representing the particle-particle correlations.
Staying within the Hartree approximation for the mean-field, the 
relativistic Hartree-Bogoliubov equations are written as
\begin{equation}
\left(\begin{array}{cc} \hat{h}_D-m-\lambda & \hat{\Delta} 
\\ -\hat{\Delta}^* & -\hat{h}_D^*+m+\lambda \end{array}\right)
\left(\begin{array}{r} U \\ V\end{array}\right)_k~=~
E_k\,\left(\begin{array}{r} U \\ V\end{array}\right)_k,
\label{RHB} 
\end{equation}
where $\hat{h}_D$ is the single-particle Dirac Hamiltonian with 
quasi-particle energies $E_k$ and the chemical potential $\lambda$.
Herein, the pairing field 
\begin{equation}
\hat{\Delta}_{ab}({\bf r},{\bf r'})~=~\frac{1}{2}\sum_{cd} V_{abcd}({\bf
 r},{\bf r'}) \kappa_{cd}({\bf r},{\bf r'})
\label{pair}
\end{equation}
is the sum over matrix elements $V_{abcd}({\bf r}, {\bf r'})$ of a two-body 
pairing interaction and the corresponding pairing tensor is defined by
\begin{equation}
\kappa_{cd}({\bf r},{\bf r'}) = \sum_{E_k > 0}U^*_{ck}({\bf r})V_{dk}({\bf
  r'}).
\label{tensor}
\end{equation}
The RHB equations are solved self-consistently in order to obtain 
eigensolutions and eigenvalues in a single quasi-particle basis. This is 
transformed to the canonical basis to obtain the desired observables.
We have used the Gogny force D1S \cite{Berger.84} for the pairing channel 
as in other works \cite{Lala.01}. We have adjusted the strength of the 
pairing force in order to get a good agreement of the calculated binding 
energies with the experimental values for the chain of Sn isotopes. 
The superfluid Sn nuclei provide a best testing ground 
for calibration of the pairing strength. 
Consequently, the force NL-SV1 describes the ground-state 
binding energies over a large range of Sn isotopes very well.

Deformed RHB calculations with the pairing channel thus calibrated have been
performed for the Kr isotopes using an expansion of Dirac spinors and mesons
fields into 20 shells of an axially deformed oscillator potential. In each
case, minimizations have been sought in the oblate and prolate regions
of deformation. It is noticed that there is a shape-coexistence between 
an oblate and a prolate shape for several isotopes. The RHB calculations with
NL-SV1 provide a good description of the ground-state binding energies of
Kr isotopes with a few divergences within 0.20\%.  

The results obtained on the isotope shifts with NL-SV1 with the 
deformed RHB are shown in Fig.~1. The RHB results 
from $N=48$ towards $N=40$ show an improvement
over the BCS values including that for $N=44$. The most interesting outcome 
of these calculations is the downward trend that arises for the isotopes 
below $N=40$, a picture that contrasts sharply with the BCS pairing.
NL-SV1 overestimates the isotope shift (charge radius) of $^{72}$Kr 
($N=36$) slightly. This is due perhaps to a slightly higher 
deformation ($\beta_2 = -0.34$) obtained theoretically than is 
the case. There is no datum on $^{70}$Kr. We predict a negative 
isotope shift for this nucleus. Results obtained with the set NL3 
also reproduce the downward trend. The NL3 results, 
however, overestimate the experimental data. 

The case of $^{72}$Kr is noteworthy. It is a $N=Z$ nucleus that participates
significantly in the rp-process. The Nilsson splitting of $j$-levels 
especially near the Fermi surface carves out a major shell gap both 
at $N=36$ and $Z=36$. Our results with the deformed RHB calculations
show that pairing correlations vanish completely for protons and neutrons
alike. This renders the nucleus $^{72}$Kr as doubly magic in the 
deformed space. Thus, in spite of being close to the proton drip line, 
the absence of smearing engenders the decrease in the charge 
radius in stark contrast to the BCS result (see Fig.~1). 

The isotopes $^{70}$Kr and $^{74}$Kr emerge with a major shell gap with
vanishing pairing at the respective neutron numbers $N=34$ and $N=36$ 
in the ground state, thus exhibiting a neutron magicity only. 
This is, to a large extent, responsible for suppressing the charge 
radius of these isotopes. 

With the advent of the double magicity in $^{72}$Kr, we have also 
explored other $N=Z$ rp-process nuclei. Results of the deformed 
RHB calculations with NL-SV1 show that not only
$^{72}$Kr, but also the other $N=Z$ nuclei such as $^{68}$Se, $^{76}$Sr and 
$^{80}$Zr exhibit larger shell gaps both at the respective neutron and
proton numbers. In these nuclei pairing correlations vanish completely for 
neutrons and  protons, which characterizes these nuclei also as doubly 
magic. Thus, the nuclei $^{68}$Se, $^{72}$Kr, 
$^{76}$Sr and $^{80}$Zr with $N=Z=$ 34, 36, 38, and 40, 
respectively, are endowed with a double magicity in the deformed space. 
It may be remarked that though deformed shell gaps at some of these N and 
Z numbers have been indicated in the literature (e.g. ref. \cite{Naza.85}), 
these shell gaps have never been shown to qualify for a magicity. 

\begin{figure}[h*]
\resizebox{0.63\textwidth}{!}{%
  \rotatebox{0}{\includegraphics{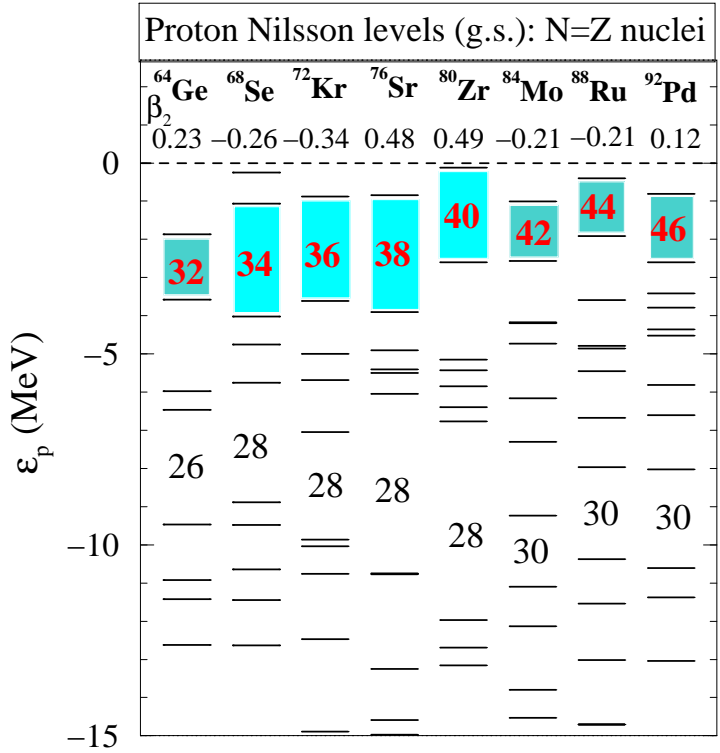}}
}
\caption{Proton single-particle levels for the ground-state of $N=Z$ nuclei 
obtained with deformed RHB approach using the force NL-SV1. Nuclei with 
the vanishing pairing correlations (proton magicity) are highlighted 
with a larger shell gap in columns 2-5.
\label{fig2}}
\end{figure}
The ensuing single-particle structure of $N=Z$ nuclei is shown in Figs.~2 and
3, where the proton and neutron Nilsson single-particle levels obtained 
with the deformed RHB approach using NL-SV1 are displayed.
The larger shell gaps at $N$ and $Z$ = 34, 36, 38, and 40
can be seen conspicuously for the aforesaid $N=Z$ nuclei.  
The corresponding $\beta_2$ for the ground-state
is indicated in the upper part of the figures. It is noteworthy that
$^{76}$Sr and $^{80}$Zr exhibit a large prolate deformation in 
the ground state. This is consistent with the experimental values deduced
\cite{Lister.87,Nacher.04}. 
\begin{figure}
\resizebox{0.63\textwidth}{!}{%
  \rotatebox{0}{\includegraphics{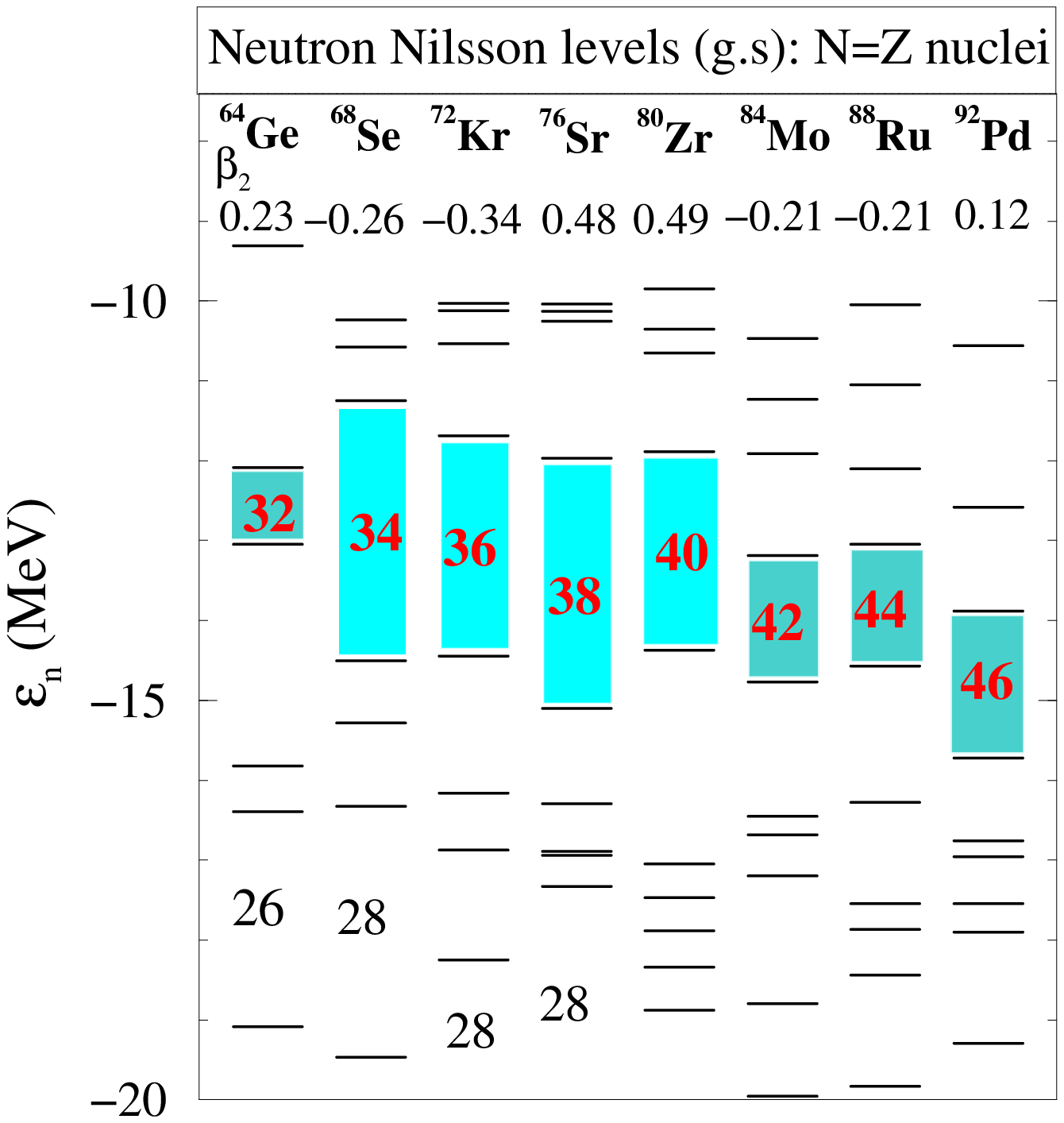}}
}
\caption{Neutron single-particle levels for the ground-state of $N=Z$ nuclei 
obtained with deformed RHB approach using the force NL-SV1. Nuclei with the 
vanishing pairing correlations (neutron magicity) are highlighted in 
columns 2-5.
\label{fig3}}
\end{figure}

The case of $^{68}$Se deserves a mention. This nucleus has been 
shown  experimentally  to be a waiting-point nucleus 
beyond doubt \cite{Gomez.08}. Our results show that this nucleus 
has a second minimum with a prolate shape within 
0.5 MeV of the oblate ground-state, thus exhibiting a shape-coexistence.
Interestingly, the single-particle structure of the prolate state also
exhibits a doubly magic character. In comparison, $^{64}$Ge ($N=Z=32$)
shows shell gaps in neutrons and protons which can not be construed
as a magic number. The corresponding pairing energy for neutrons and
protons is non-zero albeit significantly reduced as compared to a normal
superfluid nucleus. Other $N=Z$ nuclei such 
as $^{84}$Mo, $^{88}$Ru, and $^{92}$Pd also show a 
well-deformed shape in the ground state and a
shell gap both in protons and neutrons near the Fermi surface similar
in magnitude to that in $^{64}$Ge with non-zero and yet significantly 
reduced pairing correlations. These nuclei are also in contention for
being waiting-point nuclei experimentally \cite{Schatz.98}. However, our 
results do not lend a doubly-magic character to these nuclei.
\begin{figure}
\resizebox{0.70\textwidth}{!}{%
  \rotatebox{0}{\includegraphics{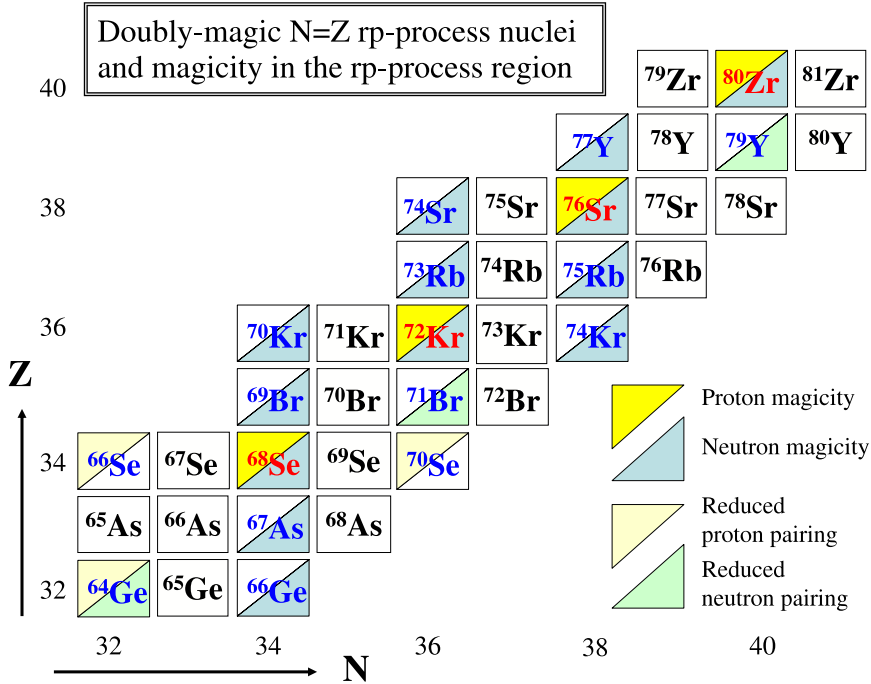}}
}
\caption{The chart of $N=Z$ nuclei exhibiting double magicity along the
rp-process path. Nuclides exhibiting neutron magicity can be seen in 
the columns with $N=$ 34, 36, 38 and 40. A few cases with the reduced 
pairing for neutrons and protons are also shown.
\label{fig4}}
\end{figure}

We have examined the single-particle structure of other nuclei in the
vicinity of the rp-process path from $N=Z=$ 32-40. Deformed RHB calculations 
have been performed for sets of nuclei as shown in Fig.~4. 
As is prevalent in this region, ground state of nuclei 
leads to a significant deformation for most of the nuclei explored. Some
cases also show a shape-coexistence between an oblate and a prolate shape.
Details of this study will be presented elsewhere. In Fig.~4 we show the
nuclei which exhibit a neutron and/or a proton magicity. The four $N=Z$
nuclei $^{68}$Se, $^{72}$Kr, $^{76}$Sr, and $^{80}$Zr with double
magicity stand out conspicuously.

We note that there is a preponderance of neutron magicity at several
neutron numbers. This includes isotones with $N=$ 34, 36, 38 and to
a limited extent with $N=40$. Surprisingly, a large number of nuclei
in this region are amenable to a larger neutron shell 
gap at the Fermi surface in the Nilsson scheme with the vanishing pairing 
correlations. A few cases with significantly reduced neutron pairing 
correlations at $N=$ 32, 36 and 40 with the pairing energy of $\sim-1$ 
to $-2$ MeV are also shown in Fig.~4. In contrast, there are no cases of 
proton magicity found other than those of four $N=Z$ doubly magic nuclei. 
Comparatively, the isotopes of $^{64}$Ge, $^{66}$Ge ($Z=32$) and 
$^{70}$Se ($Z=34$) exhibit proton shell gaps with non-zero yet significantly 
reduced proton pairing energy. Thus, proton magicity in this region is 
rather subdued as compared to neutrons.

It is interesting to note that peaks in abundances of masses at 
$A=$ 64, 68, 72, 76, and 80 were obtained in rp-process calculations 
\cite{Schatz.01} in an x-ray burst, testifying to an important role played
by the $N=Z$ = 32-40 nuclei. A double magicity of nuclei found in our work
will reinforce the waiting-point character to its namesake. It would be
interesting to investigate as to how this attribute would influence the
$\beta$-decay half-lives of nuclei and what would its effect on x-ray flux 
emanating from x-ray binaries be?

In conclusion, on the basis of our investigation of the experimental 
isotope shifts of Kr nuclei near the proton drip-line within the framework 
of the relativistic Hartree-Bogoliubov approach, we have found that 
$N=Z$ rp-process waiting-point nuclei $^{68}$Se, $^{72}$Kr, $^{76}$Sr, 
and $^{80}$Zr exhibit double magicity in the deformed space. It is also
shown that there is a preponderance of nuclides  exhibiting a neutron
magicity at $N=$ 34, 36, 38 and 40 in this region. 

One of the authors (MMS) thanks Hendrik Schatz for useful discussions.
This work is supported in part by Project SP04/04 of the Research 
Administration, Kuwait University.

\end{document}